\def\rfr#1{Eq. (\ref{#1})}

\def\dert#1#2{\frac{{{d}}{#1}}{{{d}}{#2}}}              

\def\bar{\begin{eqnarray}}
\def\ear{\end{eqnarray}}

\def\eqi{\begin{equation}}
\def\eqf{\end{equation}}
\def\eqia{\begin{eqnarray}}
\def\eqfa{\end{eqnarray}}
\def\rp#1#2{{#1\over#2}}

\def\lb#1{\label{#1}}

\def\oc2{$\mathcal{O}(c^{-2})$}

\documentclass[11pt]{article}
\usepackage{amsmath,amsthm,amscd,amssymb}
\usepackage{latexsym}
\usepackage{graphicx,epsfig,lscape,rotating}

\begin{document}

\noindent{\bf \LARGE{On the use of Ajisai and Jason-1 satellites
for tests of General Relativity }}
\\
\\
\\
{L. Iorio,  }\\
{\it Viale Unit$\grave{a}$ di Italia 68, 70125\\Bari, Italy
\\tel./fax 0039 080 5443144
\\e-mail: lorenzo.iorio@libero.it}

\begin{abstract}
In this paper we analyze in detail some aspects of the proposed
use of Ajisai and Jason-1, together with the LAGEOS satellites, to
measure the general relativistic Lense-Thirring effect in the
gravitational field of the Earth.  A linear combination of the
nodes of such satellites
is the proposed observable. The systematic error due to the
mismodelling in the uncancelled even zonal harmonics would be
$\sim 1\%$ according to the latest present-day CHAMP/GRACE-based
Earth gravity models. In regard to the non-gravitational
perturbations especially affecting Jason-1, only relatively
high-frequency harmonic perturbations should occur: neither
semisecular nor secular bias of non-gravitational origin should
affect the proposed combination: their maximum impact is evaluated
to $\sim 4\%$ over 2 years. Our estimation of the root-sum-square
total error is about 4-5$\%$ over at least 3 years of data
analysis required to average out the uncancelled tidal
perturbations.
\end{abstract}


\section{Introduction}
The most recent and relatively accurate test of the general
relativistic gravitomagnetic Lense-Thirring effect on the orbit of
a test particle (Lense and Thirring 1918; Barker and O'Connell
1974; Cugusi and Proverbio 1978; Soffel 1989; Ashby and Allison
1993; Iorio 2001) in the gravitational field of the
Earth\footnote{A more precise  (6$\%$ on average) test of the
Lense-Thirring effect was recently reported by Iorio (2006a) in
the gravitational field of Mars. } was performed by Ciufolini and
Pavlis (2004), who analyzed the laser data of the LAGEOS and
LAGEOS II satellites according to a suitable combination of the
residuals of their nodes proposed in (Ries et al. 2003a; 2003b;
Iorio and Morea 2004) \eqi\delta\dot\Omega^{\rm LAGEOS}+
c_1\delta\dot\Omega^{\rm LAGEOS\ II }\sim 48.1.\lb{iorform}\eqf

Let us briefly recall the linear combination approach from which
\rfr{iorform} originates. The combinations are obtained by
explicitly writing down the expressions of the residuals of $N$
orbital elements (the nodes of different satellites in our case)
in terms of the classical secular precessions induced by the
mismodelled part of $N-1$ even zonal harmonic coefficient $\delta
J_{\ell},\ \ell=2,4,...$ of the multipolar expansion of the
terrestrial gravitational potential (see also Section
\ref{gravzon} and \rfr{prc} for the meaning of the coefficients
$\dot\Omega_{.\ell}$) and the Lense-Thirring effect
$\dot\Omega_{\rm LT}$ considered as an entirely unmodelled feature
of motion
\eqi\delta\dot\Omega^{(i)}=\sum_{\ell=2}^{2(N-1)}\dot\Omega^{(i)}_{.\ell}\delta
J_{\ell}+\dot\Omega^{(i)}_{\rm LT}\mu_{\rm LT},\
i=1,2...N,\lb{syst}\eqf
 and solving the resulting algebraic non-homogeneous linear system of $N$ equations in $N$ unknowns
 of \rfr{syst} with respect to the scaling parameter $\mu_{\rm
LT}$ which is 1 in the Einsteinian theory and 0 in Newtonian
mechanics. The obtained coefficients weighing the satellites'
orbital elements depend on their semimajor axes $a$,
eccentricities $e$ and inclinations $i$: they allow to cancel out
the impact of the $N-1$ even zonal harmonics considered. In
\rfr{iorform} the  value of the secular trend predicted by the
General Relativity Theory is 48.1 milliarcseconds per year (mas
yr$^{-1}$), and $c_1=0.546$. The coefficient $c_1$
makes the combination of \rfr{iorform} insensitive to the biasing
action of only the first even zonal $J_2$  and its temporal
variations.

The other even zonal harmonics $J_{\ell\geq 4}$, along with their
secular variations $\dot J_{\ell\geq 4}$, do affect \rfr{iorform}
inducing a systematic error in the measurement of the
Lense-Thirring effect, whose correct and reliable evaluation is of
crucial importance for the reliability of such an important test
of fundamental physics. Ciufolini and Pavlis (2004), who used the
GRACE-only Earth gravity model EIGEN-GRACE02S (Reigber et al
2005a), claimed a total error of $5\%$ at 1-sigma and $10\%$ at
3-sigma. Such estimates were criticized by Iorio (2005; 2006b) for
various reasons. His evaluations, based on the analysis of
different gravity model solutions and on the impact of the secular
variations of the uncancelled even zonals, point toward a more
conservative $\sim 20\%$ total error at 1-sigma.

The major drawbacks of the combination of \rfr{iorform} are as
follows
\begin{itemize}
  \item It is mainly affected  by the low-degree even zonal
  harmonics $J_4, J_6$.
  The combination of  \rfr{iorform} is practically insensitive to the even
  zonal harmonics of degree higher than $\ell=12-14$ in the sense that the error induced by the
  uncancelled zonals does not change
if the terms of degree higher than $\ell=12-14$ are neglected in
the calculation, as fully explained in Section \ref{gravzon}.
  Unfortunately, the major improvements from the present-day and forthcoming
  GRACE models are mainly expected just for the medium-high degree
  even zonal harmonics which do not affect \rfr{iorform}.
  Instead, the low-degree even zonals should not experience
  notable improvements, as showed by the most recent long-term models like
  EIGEN-CG01C (Reigber et al. 2006), EIGEN-CG03C (F\"{o}rste et al. 2005), EIGEN-GRACE02S, GGM02S
  (Tapley et al. 2005). Moreover, the part
  of the systematic error due to them is still rather
  model-dependent ranging from $\sim 4\%$ to $\sim 9\%$.
  \item Another source of aliasing for the combination of \rfr{iorform} is represented by the
  secular variations  $\dot J_4$ and $\dot J_6$ whose  signal
  grows
  quadratically in time. Their bias on the measurement of the Lense-Thirring effect with
  the combination of \rfr{iorform} was evaluated to be of the order of $\sim 10\%$ (Iorio 2005).
  They are, at present, known with modest
  accuracy and there are few hopes that the situation could become
  more favorable in the near future. Moreover, also interannual
  variations of $J_4$ and $J_6$ may turn out to occur
\end{itemize}
Thus, it seems unlikely that relevant improvements in the
reliability and accuracy of the tests conducted with the adopted
node-node combination of the LAGEOS satellites will occur in the
foreseeable future.

In Iorio and Doornbos (2005) the following combination
\eqi\delta\dot\Omega^{\rm LAGEOS}+k_1\delta\dot\Omega^{\rm LAGEOS\
II}+k_2\delta\dot\Omega^{\rm Ajisai}+k_3\delta\dot\Omega^{\rm
Jason-1}=\mu_{\rm LT}49.5,\lb{jason}\eqf with \eqi k_1=0.347,\
k_2=-0.005,\ k_3=0.068,\lb{jasoncoef}\eqf was designed: it comes
from \rfr{syst} applied to the nodes of LAGEOS, LAGEOS II, Jason-1
and Ajisai. A similar proposal was put forth by Vespe and
Rutigliano (2005): however, the less accurate CHAMP-only Earth
gravity model EIGEN3p (Reigber et al. 2005b) was used in that
exhaustive analysis.
Such a
combination involves the nodes of the geodetic Ajisai satellite
and of the radar altimeter Jason-1 satellite. Their orbital
parameters, together with those of the LAGEOS satellites, are
listed in Table \ref{sat_par}. The combination of \rfr{jason}
allows cancellation of the first three even zonal harmonics
$J_2,J_4, J_6$ along with their temporal variations. The resulting
systematic error of gravitational origin is of the order of $\sim
1\%$. The practical implementation of the proposed test would
consist in the following three stages
\begin{itemize}
  \item The best possible nodes from independent arcs of data (for example,
  weekly) will be assembled as a time-series for the four
  satellites
  \item Correspondingly, an integrated long-term node time-series will be
  constructed for each satellite with the best available dynamical models
  not using such force models derived empirically from the same
  tracking data determining the observed nodes (otherwise, also the Lense-Thirring effect would be removed.
  See Section \ref{cazzo})
  \item From such two time-series a residual time-series will be
  built up for each satellite, combined according to \rfr{jason} and analyzed for both secular and periodic terms; the
  secular component will be used to extract the Lense-Thirring
  effect
\end{itemize}

The goal of the present paper is to analyze in detail some
important critical aspects of the use of such a combination. They
are
\begin{itemize}
  \item The impact of the higher degree even zonal harmonics
  introduced by the lower orbiting satellites Ajisai and Jason-1
  \item The impact of the realistically obtainable
  accuracy of a truly dynamical orbital reconstruction for Ajisai and
  Jason-1
  \item The impact of the atmospheric drag and of the other
  non-gravitational perturbations on Ajisai and, especially,
  Jason-1
\end{itemize}

\section{Systematic error due to even zonal
harmonics}\lb{gravzon} The even zonal harmonics $J_{\ell},\
\ell=2,4,6,...$ of the Newtonian multipolar expansion of the
Earth's gravitational potential induce on the node of an
artificial satellite a classical secular precession which can be
cast in the form \eqi\dot\Omega^{\rm class }=\sum_{\ell\geq
2}\dot\Omega_{.\ell}J_{\ell}\lb{prc}.\eqf The coefficients
$\dot\Omega_{.\ell}$ depend on the Earth's $GM$ and mean
equatorial radius $R$, and on the semimajor axis, the eccentricity
and the inclination of the satellite. They were analytically
calculated up to degree $\ell=20$ in Iorio (2003) and their
numerical values, in mas yr$^{-1}$, for LAGEOS, LAGEOS II, Ajisai
and Jason-1 can be found in Table \ref{prec_coef}. The
coefficients $c$ and $k$ of the combinations of \rfr{iorform} and
\rfr{jason} are built up with $\dot\Omega_{.\ell}$.

The precessions of \rfr{prc} are much larger than the
Lense-Thirring rates. This is the reason why the combinations of
\rfr{iorform} and \rfr{jason} are, by construction, designed in
order to cancel out the precessions induced by the first
low-degree even zonals. This approach was proposed for the first
time by Ciufolini (1996) with a combination involving the nodes of
LAGEOS and LAGEOS II and the perigee of LAGEOS II.

One of the major objections about the combination also involving
Ajisai and Jason-1 is that such satellites, which orbit at much
lower altitudes with respect to LAGEOS and LAGEOS II, (Table
\ref{sat_par}), would introduce much more even zonals in the
systematic error of gravitational origin than the node-node
LAGEOS-LAGEOS II combination. Indeed, the classical precessions
depend on the satellite's semimajor axis as
\eqi\dot\Omega_{.\ell}\propto
a^{-\left(\rp{3}{2}+\ell\right)}.\eqf In fact, this criticism
would better fit the case of the other existing low-orbit geodetic
satellites like, e.g., Starlette and Stella. Indeed, Iorio (2006c)
showed that, according to EIGEN-CG03C, a combination involving
such spacecraft is not yet competitive with other combinations
just because of the systematic bias due to the even zonals. In the
case of the combination of \rfr{jason}, it turns out that only
about the first ten even zonals are relevant for a satisfactorily
estimate of the systematic error of gravitational origin. Indeed,
it can be evaluated as \eqi \delta\mu_{\rm LT}\leq\sum_{\ell\geq
8}\left|\left(\dot\Omega_{.\ell}^{\rm
LAGEOS}+k_1\dot\Omega_{.\ell}^{\rm LAGEOS\
II}+k_2\dot\Omega_{.\ell}^{\rm Ajisai}+k_3\dot\Omega_{.\ell}^{\rm
Jason-1}\right)\right|\delta J_{\ell},\lb{erro}\eqf where $\delta
J_{\ell}$ are the errors in the even zonal harmonics according to
a given Earth gravity models. They can be found in Table
\ref{geopot_models} for EIGEN-CG03C, EIGEN-CG01C, EIGEN-GRACE02S
and GGM02S.

Note that \rfr{erro} yields a conservative upper bound of the bias
induced by the mismodelling in the even zonal harmonics. The
individual terms of \rfr{erro}, calculated for the various gravity
solutions considered here, are listed in Table \ref{comb_err}. It
can be noted that the resulting total error $\delta\mu_{\rm LT}$,
which is of the order of $\sim 1\%$ of the Lense-Thirring effect
for the latest models combining data from CHAMP, GRACE and
ground-based measurements, does not significantly change if the
terms of degree higher than $\ell=20$ are not included in the
calculation.

Another important feature is that such an error is much less
model-dependent than that of the combination of \rfr{iorform};
moreover, it is likely that the forthcoming gravity models based
on CHAMP and GRACE will further ameliorate the situation because
they should especially improve the medium-high degree even zonal
harmonics to which the combination of \rfr{jason} is mainly
sensitive. Other distinctive features of \rfr{jason} are that the
secular variations $\dot J_2,\ \dot J_4,\ \dot J_6 $ do not affect
it by construction and no long-period harmonic perturbations of
tidal origin would corrupt the measurement of the Lense-Thirring
effect. Indeed, the most powerful uncancelled tidal perturbation
is due to the $\ell=2,m=1$ component of the solar $K_1$ tide whose
period is equal to the satellite's node period: the periods of the
nodes of LAGEOS, LAGEOS II, Ajisai and Jason amount to 2.84,
-1.55, -0.32 and -0.47 years, respectively.

In conclusion, the systematic error of gravitational origin of the
combination of \rfr{jason} can safely be evaluated as $\sim 1\%$:
the forthcoming Earth gravity models from CHAMP and, especially,
GRACE  should be able to reduce such error below the $1\%$ level.
\section{The orbital reconstruction accuracy}
Another possible criticism about the use of Ajisai and Jason-1 for
measuring the Lense-Thirring effect is the following. While the
cm-level accuracy in reconstructing the orbits of the LAGEOS
satellites is based on a truly dynamical, very extensive and
accurate modelling of the various accelerations of
non-gravitational origin affecting them, it would not be so for
Ajisai and, especially, Jason-1. Indeed, too many empirical
accelerations which could sweep out also the Lense-Thirring effect
of interest  are used to reach the cm level for Ajisai and Jason-1
(Lutchke et al. 2003). Without resorting to such a reduced-dynamic
approach, the genuine obtainable orbit accuracy would be worst.

Let us assume, very conservatively, that the root--mean--square
(RMS) of the recovered orbits of Ajisai and Jason-1, amount to 1 m
over, say, 1 year. Then, the error in the nodal rates can be
quantified as 26.2 mas and 26.7 mas for Ajisai and Jason-1,
respectively. Thus, their impact on the combination \rfr{jason}
would amount to 1.6 mas, i.e. about $3\%$ of the Lense-Thirring
effect over 1 year. In view of the fact that the temporal interval
of the analysis should cover some years and that a more realistic
estimate of the orbital accuracy amounts to some tens of cm, it
can be concluded that the impact of the orbital reconstruction
errors on the combination of \rfr{jason} is at the few percent
level. However, it must be stressed that this is a very
pessimistic evaluation because, even at this level for Jason-1 or
Ajisai such independent errors for weekly orbits would result in
totally negligible secular rate errors on fitting the weekly
time-series over a number of years.
\section{The impact of the non-gravitational perturbations}
The impact of the non-gravitational perturbations on the
LAGEOS-type satellites has been the subject of numerous recent
papers (Ries et al. 1989; Lucchesi 2001; 2002; 2003; 2004;
Lucchesi et al 2004); according to the most recent works, it turns
out that the systematic bias induced by them on the combination of
\rfr{jason} through LAGEOS and LAGEOS II would be of the order of
$1\%$.

Undoubtedly the major concern about the use of the combination of
\rfr{jason}, the non--gravitational perturbations affect Ajisai
and, especially, Jason-1 much more severely than LAGEOS and LAGEOS
II. Indeed, the area-to-mass ratios, to which such kind of
perturbing effects are proportional, of Ajisai and Jason-1 are
larger than those of LAGEOS and LAGEOS II by one or two orders of
magnitude. They are listed in Table \ref{sat_par}. Moreover,
Jason-1 is not spherical in shape, is endowed with steering solar
panels and is regularly affected by orbital maneuvers due to its
primary altimetric and oceanographic tasks. For example, the
pointing of the solar panels to the Sun is not perfectly normal to
the solar direction, so that there may be small systematic
cross-track forces whose impact is difficult to be reliably
assessed. There are also many complicated reflective and emissive
surfaces on the spacecraft bus. As the solar panels produce power
for the onboard instruments and heaters, some part of it is
routinely dumped into space by heat radiators on the side of the
spacecraft, in particular when the batteries are fully re-charged.
Since the heat cannot be dumped to the side of the spacecraft
exposed to the Sun, part of the yaw-steering algorithm is intended
to keep this side away from solar exposure. Such heat re-radiation
acceleration tends to have a component in the cross-track
direction which probably has some orbital period dependence.
Another point to be considered is that the radiators are typically
only on one side of Jason-1, so that the satellite performs a
`yaw-flip' each time the orbital plane passes through the solar
direction to keep that side away from the Sun. This fact might
yield to a non-symmetric pattern of the resulting accelerations.
As a consequence, an entirely reliable and accurate modelling of
the perturbations of non--gravitational origin acting on it is not
an easy task.

Nonetheless, in the next Sections we will show that the situation
for the combination of \rfr{jason} is less unfavorable than it
could seem at a first sight, provided that some simplifying
assumptions are made.
\subsection{The non-gravitational accelerations on Ajisai}
Ajisai is a spherical geodetic satellite launched in 1986. It is a
hollow sphere covered with 1436 corner cube reflectors (CCRs) for
SLR and 318 mirrors to reflect sunlight. Its diameter is 2.15 m,
contrary to LAGEOS which has a diameter of 60 cm. Its mass is 685
kg, while LAGEOS mass is 406 kg. Then, the Ajisai's area-to-mass
ratio $S/M$, which the non--conservative accelerations are
proportional to, is larger than that of the LAGEOS satellites by
almost one order of magnitude resulting in a higher sensitivity to
surface forces. However, we will show that their impact on the
proposed combination \rfr{jason} should be less than 1$\%$.
\subsubsection{The atmospheric drag}
An important non--conservative force  affecting the orbits of
low-Earth  satellites is the atmospheric drag. Its acceleration
can be written as \eqi
\textbf{\textit{a}}_{D}=-\rp{1}{2}C_D\left(\rp{S}{M}\right)\rho V
\textbf{\textit{V}},\eqf where $C_D$ is a dimensionless drag
coefficient close to 2, $\rho$ is the atmospheric density and
\textbf{\textit{V}} is the velocity of the satellite relative to
the atmosphere (called ambient velocity). Let us write
\textbf{\textit{V}}=\textbf{\textit{v}}$-{\boldsymbol{
\sigma}}\times$ \textbf{\textit{r}} where \textbf{\textit{v}} is
the satellite's velocity in an inertial frame. If the atmosphere
corotates with the Earth ${\boldsymbol \sigma}$ is the Earth's
angular velocity vector
${\boldsymbol\omega}_{\oplus}=\omega_{\oplus}$\textbf{\textit{k}},
where \textbf{\textit{k}} is a unit vector. However, it must be
considered that there is a 20$\%$ uncertainty in the corotation of
the Earth's atmosphere at the Ajisai's altitude. Indeed, it is
believed that the atmosphere rotates slightly faster than the
Earth at some
 altitudes with a 10-20$\%$ uncertainty. We will
then assume ${\boldsymbol
\sigma}=\omega_{\oplus}(1+\xi)$\textbf{\textit{k}}, with $\xi
=0.2$ in order to account for this effect.

Regarding the impact of a perturbing acceleration on the orbital
motion, the Gaussian perturbative equation for the nodal rate is
\eqi\dert{\Omega}{t}=\frac{1}{na\sqrt{1-e^2}\sin
i}A_N\left(\frac{r}{a}\right)\sin u,\lb{nodogaus}\eqf where
$n=\sqrt{GM/a^3}$ is the Keplerian mean motion, $A_N$ is the
out--of--plane component of the perturbing acceleration and
$u=\omega+f$ is the satellite's argument of latitude.

The out--of--plane acceleration induced by the atmospheric drag
can be written as (Abd El-Salam and Sehnal 2004) \eqi A_N^{(\rm
atm )}=-\rp{1}{2}K_D\sigma\rho vr\sin i\cos u,\lb{atmo}\eqf with
\eqi K_D=C_D\left(\rp{S}{M}\right)\sqrt{k_R}\eqf and \eqi
k_R=1-\rp{2\sigma h\cos i}{v^2}+\left(\rp{\sigma r
\cos\delta}{v}\right)^2 . \eqf The quantities $h$ and $\delta$ are
the orbital angular momentum per unit mass and the satellite's
declination, respectively. By inserting \rfr{atmo} in
\rfr{nodogaus} and evaluating it on an unperturbed Keplerian
ellipse it can be obtained \eqi \rp{d\Omega}{dt}^{(\rm atm)
}\propto -\rp{1}{2}K_D \rho(1-e^2)\sigma a.\lb{atmoden}\eqf It
must be pointed out that the density of the atmosphere $\rho$ has
many irregular and complex variations both in position and time.
It is largely affected by solar activity and by the heating or
cooling of the atmosphere. Moreover, it is not actually
spherically symmetric but tends to be oblate. A very cumbersome
analytic expansion of $\rho$ based on the TD88 model can be found
in (Abd El-Salam and Sehnal 2004). In order to get an order of
magnitude estimate we will consider a typical value $\rho=1\times
10^{-18} $ g cm$^{-3}$ at Ajisai altitude (Sengoku et al. 1996).
By assuming $C_D=2.5$ \rfr{atmoden} yields a nominal amplitude of
25 mas yr$^{-1}$ for the atmosphere corotation case and 5 mas
yr$^{-1}$ for the 20$\%$ departure from exact corotation. The
impact of such an effect on \rfr{jason} would be $3\times
10^{-3}$.
\subsubsection{The thermal and radiative forces} The action of the
thermal forces due to the interaction of solar and terrestrial
electromagnetic radiation with the complex physical structure of
Ajisai has been investigated in Sengoku et al. (1996). The
temperature asymmetry on Ajisai caused by the infrared radiation
of the Earth produces a force along the satellite spin axis
direction called the Yarkovsky-Rubincam effect. This thermal
thrust produces secular perturbations in the orbital elements, but
no long-periodic perturbations exist if the spin axis of Ajisai is
aligned with the Earth's rotation axis. In fact, the spin axis was
set parallel to the Earth rotation axis at orbit insertion. The
analogous solar heating (Yarkovsky-Schach effect) is smaller than
the terrestrial heating. A nominal secular nodal rate of 15 mas
yr$^{-1}$ due to the Earth heating has been found. It would affect
\rfr{jason} at a $1.5\times 10^{-3}$ level.

The effect of the direct solar radiation pressure on Ajisai has
been studied in Sengoku et al. (1995). For an axially symmetric,
but not spherically symmetric, satellite like Ajisai, there is a
component of the radiation pressure acceleration directed along
the sun-satellite direction $a^{(\rm iso)}_{\odot}$ and another
smaller component perpendicular to the sun-satellite direction. By
assuming for the isotropic reflectivity coefficient its maximum
value $C_R=1.035$, the radiation pressure acceleration $a^{(\rm
iso )}_{\odot}$ experienced by Ajisai amounts to $2.5\times
10^{-8}$ m s$^{-2}$. Its nominal impact on the node, proportional
to $ea^{(\rm iso )}_{\odot}/4na$, can be quantified as 5.7 mas
yr$^{-1}$; it yields a $5.7\times 10^{-4}$ relative error on
\rfr{jason}. The anisotropic component of the acceleration would
amount, at most, to 2$\%$ of the isotropic one, so that its impact
on \rfr{jason} would be totally negligible.
\subsection{The non-gravitational accelerations on Jason-1}\lb{cazzo}
As already previously noted, the complex shape, varying attitude
modes and the relatively high area--to--mass ratio of Jason-1,
suggest a more complex modelling and higher sensitivity to the
non--gravitational accelerations than in the case of the spherical
geodetic satellites. On the other hand, important limiting factors
in the non-gravitational force modelling of the spherical
satellites are attitude and temperature knowledge. These
parameters are actually very well--defined and accurately measured
(Marshall et al. 1995) on satellites such as Jason--1. A lot of
effort has already been put into the modelling of
non--gravitational accelerations for TOPEX/Poseidon (Antreasian
and Rosborough 1992, Marshall et al. 1994; Kubitschek and Born
2001), so that similar models (Berthias et al. 2002) have been
routinely implemented for Jason--1.

These so--called box--wing models, in which the satellite is
represented by eight flat panels, were developed for adequate
accuracy while requiring minimal computational resources. A recent
development is the work on much more detailed models of satellite
geometry, surface properties, eclipse conditions and the Earth's
radiation pressure environment for use in orbit processing
software (Doornbos et al. 2002; Ziebart et al. 2003). It should be
noted that such detailed models were not yet adopted in the orbit
analyses by Lutchke et al. (2003). In fact, their results were
based on the estimation of many empirical 1-cycle-per-revolution
(cpr) along-track and cross-track acceleration parameters, which
absorb all the mismodelled/unmodelled physical effects, of
gravitational and non-gravitational origin, which induce secular
and long-period changes in the orbital elements. Due to the power
of this reduced--dynamic technique, based on the dense tracking
data, further improvements in the force models become largely
irrelevant for the accuracy of the final orbit. Such improved
models remain, however, of the highest importance for the
determination of $\delta\dot\Omega^{\rm Jason-1}$.

From \rfr{nodogaus} it can be noted that, since we are interested
in the effects averaged over one orbital revolution, the impact of
every acceleration constant over such a timescale would be
averaged out. As previously noted, the major problems come from
1--cpr out--of--plane accelerations of the form $A_N=S_N\sin
u+C_N\cos u$, with $S_N$ and $C_N$ constant over one orbital
revolution.

\begin{figure}[htb]
\center{\includegraphics[width=1.00\textwidth]{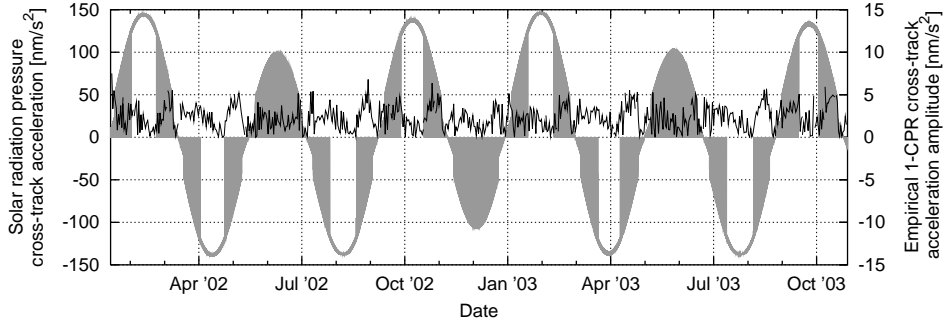}}
\caption{\label{fig:solrad}Time series of the modelled Jason--1
out--of--plane solar radiation pressure acceleration (grey) and
the estimated 1--cpr out--of--plane acceleration (black).
}
\end{figure}

We have analyzed both the output of the non-gravitational force
models described in Berthias et al. (2002) and the resulting
residual empirical 1--cpr accelerations estimated from DORIS and
SLR tracking over 24--hour intervals. Solar radiation pressure,
plotted in Figure \ref{fig:solrad}, is by far the largest
out--of--plane non--gravitational acceleration, with a maximum
amplitude of 147 nm s$^{-2}$. It is followed by Earth radiation
pressure at approximately 7 nm s$^{-2}$.  The contributions of
aerodynamic drag and the thermal imbalance force on the
cross--track component are both estimated to have a maximum of
approximately 0.5 nm s$^{-2}$. As can be seen in Figure
\ref{fig:solrad}, the cross--track solar radiation pressure
acceleration shows a sinusoidal long-term behavior, crossing zero
when the Sun--satellite vector is in the orbital plane, roughly
every 60 days. It is modulated by the long--term seasonal
variations in Sun--Earth geometry, as well as by eclipses and the
changing satellite frontal area, both of which contribute 1--cpr
variations. In fact, the shading of certain areas in Figure
\ref{fig:solrad} is due to the effect of the eclipses, which, at
once--per--orbit, occur much more frequently than can be resolved
in Figure \ref{fig:solrad}.

As mentioned before, the empirical 1--cpr accelerations absorb the
errors of almost all the unmodelled or mismodelled forces.  Now
note the systematic way in Figure \ref{fig:solrad}, in which the
empirical 1--cpr cross-track acceleration drops to values of below
1 nm~s$^{-2}$ near the end of each eclipse--free period, and has
its maximum level of 5--6 nm~s$^{-2}$ only during periods
containing eclipses.  The fact that the amplitude, but also the
phase (not shown in Figure \ref{fig:solrad}) of the 1--cpr
accelerations show a correlation with the orientation of the
orbital plane with respect to the Sun, indicates that it is for a
large part absorbing mismodelled radiation pressure accelerations.

By averaging \rfr{nodogaus} over one orbital revolution and from
the orbital parameters of Table \ref{sat_par} it turns out that a
1--cpr cross--track acceleration would induce a secular rate on
the node of Jason proportional to $7.6 \times 10^{-5}$ s
m$^{-1}\times S_N$ m~s$^{-2}$. This figure must be multiplied by
the combination coefficient $k_3$.  By using the average value of
the empirical 1--cpr acceleration from the above analysis
$S_N\approx$ 2.3 nm~s$^{-2}$ as an estimate for the mismodelled
non--gravitational forces, it can be argued that the impact on
$k_3\delta\dot\Omega^{\rm Jason-1}$ would amount to 77.4 mas
yr$^{-1}$.

However, it must be pointed out that our assumed value of $S_N$
can be improved by adopting the aforementioned more detailed force
models or by tuning the radiation pressure models using tracking
data. In addition, it must be pointed out that $S_N$ experiences
long--term variations mainly induced by the orientation of the
orbital plane with respect to the Sun, and the related variations
in satellite attitude. For Jason--1 such a periodicity amounts to
approximately 120 days (the $\beta^{'}$ cycle). Let us, now,
evaluate what would be the impact of such a long-periodic
perturbation on our proposed measurement of the Lense--Thirring
effect. Let us write, e.g., a sinusoidal law for the
long--periodic component of the weighted nodal rate of Jason--1
\eqi k_3\frac{d\Omega}{dt}=(77.4\ {\rm mas\ yr^{-1}})\times
\cos\left[2\pi\left(\rp{t}{P_{\rm \beta^{'}
}}\right)\right];\lb{booo}\eqf then, if we integrate \rfr{booo}
over a certain time span $T_{\rm obs}$ we get \eqi
k_3\Delta\Omega=\left(\rp{P_{\beta^{'}}}{2\pi}\right)(77.4\ {\rm
mas\ yr^{-1}} )\times\sin\left[2\pi\left(\rp{T_{\rm obs
}}{P_{\beta^{'}}}\right)\right].\eqf Then, the amplitude of the
shift due to the weighted node of Jason--1, by assuming
$P_{\beta^{'}}\cong 120$ days, would amount to \eqi k_3
\Delta\Omega\leq 4\ {\rm mas}.\eqf The maximum value would be
obtained  for \eqi \rp{T_{\rm obs}}{P_{\beta^{'}}}=\rp{j}{4},\
j=1,3,5,...\cong 30, 90, 150,...\ {\rm days}. \eqf So, the impact
on the proposed measurement of the Lense--Thirring effect would
amount to \eqi \left.\rp{\delta\mu_{\rm LT}}{\mu_{\rm
LT}}\right|_{\rm SRP}\leq\rp{(4\ {\rm mas})}{(49.5\ {\rm mas\
yr^{-1}})\times (T_{\rm obs}\ {\rm yr})};\lb{limit}\eqf for, say,
$T_{\rm obs}=2$ years \rfr{limit} yields an upper bound of  $4\%$.

Moreover, it must also be noted that it would be possible to fit
and remove such long--periodic signals from the time-series
provided that an observational time span longer than the period of
the perturbation is adopted.

\section{Conclusions}
In this paper the use of a suitable linear combination of the
nodes of LAGEOS, LAGEOS II, Ajisai and Jason-1 to measure the
Lense-Thirring effect in the gravitational field of the Earth is
examined. Below we list the major sources of errors along with our
evaluations of their impact on the proposed measurement. They are
also summarized in Table \ref{tabwag}.
\subsection{The gravitational error}
It turns out that the systematic error of gravitational origin due
to the even zonal harmonics can be presently evaluated to be $\sim
1\%$, according to the latest Earth gravity models based on the
combined data of CHAMP, GRACE and ground-based measurements. Such
an estimate is rather model-independent and will be likely further
improved when the new, forthcoming solutions for the terrestrial
gravitational potential will be available. The temporal variations
of the even zonal harmonics do not represent a major concern
because the secular and possible interannual variations of the
first three even zonal harmonics are cancelled out, by
construction, along with their static components. Moreover, the
uncancelled tidal perturbations, like the solar $K_1$ tide, vary
with relatively high frequencies, so that they could be fitted and
removed from the time-series or averaged out over an observational
time span of at least 3 years (the longest period is that of the
LAGEOS node amounting to 2.84 years).
\subsection{The measurement errors}
Our largely conservative evaluation for the measurement errors
amounts to $\sim 3\%/N$, where $N$ is the number of years of the
experiment duration, by assuming a really pessimistic 1 m error in
a truly dynamical orbit reconstruction for Ajisai and Jason-1 over
the adopted time span.
\subsection{The non-gravitational error}
In regard to the non-gravitational perturbations, which especially
affect Jason-1, it is worthwhile noting that no secular aliasing
trends should occur, but only high-frequency harmonic
perturbations. However, particular attention should be paid to an
as accurate as possible truly dynamical modelling of the
non-gravitational accelerations acting on the node of Jason-1.
Also a careful choice of the observational time span of the
analysis would be required in order to reduce the uncertainties
related to the orbital maneuvers which are mainly in plane,
although a small, unknown, part of them affects also the
out-of-plane part of the orbit. We evaluate the error due to the
non-gravitational accelerations as large as $\sim 4\%$ over 2
years.
\subsection{Final remarks}
In conclusion, the use of the proposed combination, although
undoubtedly  difficult and demanding, seems to be reasonable and
feasible; we give a total root-sum-square uncertainty of $\sim$
4-5$\%$ over at least 3 years required to average out the
uncancelled tidal perturbations. Moreover, the efforts required to
perform the outlined analysis should be rewarding not only for the
relativists' community but also for people involved in  space
geodesy, altimetry and oceanography.

\newpage
\section*{Acknowledgments}
I thank E Doornbos for Figure 1, many useful references and
important discussions and clarifications. I am also grateful to C
Wagner, E Grafarend and J Ries for their useful and critical
remarks and observations. I am also grateful to the anonymous
referee whose comments and observations greatly improved the
manuscript.

\newpage

\begin{table}
\caption{Orbital parameters, area-to-mass ratios $S/M$ and
Lense-Thirring node precessions $\dot\Omega_{\rm LT}$ of LAGEOS,
LAGEOS II, Ajisai and Jason-1. $a$ is the semimajor axis,  $e$ is
the eccentricity and $i$ is the inclination to the Earth's
equator.}\lb{sat_par}
\begin{center}
\begin{tabular}{lllll}
\noalign{\hrule height 1.5pt}

 & LAGEOS &  LAGEOS II & Ajisai & Jason-1\\

\hline

$a$ (km) & 12270 & 12163 & 7870 & 7713\\
$e$ & 0.0045 & 0.014 & 0.001 & 0.0001\\
$i$ (deg) & 110 & 52.65 & 50 & 66.04\\
$S/M$ (m$^2$ kg$^{-1}$) & $6.9\times 10^{-4}$ & $7.0\times
10^{-4}$ & $5.3\times 10^{-3}$ & $2.7\times 10^{-2}$\\
$\dot\Omega_{\rm LT}$ (mas yr$^{-1}$) & 30.7 & 31.4 & 116.2 &
123.4\\

\noalign{\hrule height 1.5pt}
\end{tabular}
\end{center}
\end{table}
\begin{sidewaystable}
\small{ \caption{Coefficients $\dot\Omega_{.\ell}$ of the
classical node even zonal precessions of LAGEOS, LAGEOS II, Ajisai
and Jason-1 up to degree $\ell=40$, in mas yr$^{-1}$.
}\label{prec_coef}
\begin{center}

\begin{tabular}{lllll} \noalign{\hrule height 1.5pt}
 $\ell$ & LAGEOS  & LAGEOS II & Ajisai & Jason-1\\

\hline

2 & $4.191586788514\times 10^{11}$ & $-7.669274920758\times 10^{11}$ & $-3.727536980291872\times 10^{12}$ & $-2.527057829772086\times 10^{12}$\\
4 & $1.544030247472\times 10^{11}$ & $-5.58637864293\times 10^{10}$ & $-1.648980015146924\times 10^{11}$ & $-1.993205109225784\times 10^{12}$\\
6 & $3.25092246054\times 10^{10}$& $4.99185703735\times 10^{10}$ & $1.549123674154874\times 10^{12}$ & $-6.138803379467100\times 10^{11}$\\
8 & $2.1343038821\times 10^{9}$ & $1.10707933989\times 10^{10}$ & $3.920255674567422\times 10^{11}$ & $3.707497244428685\times 10^{11}$\\
10 & $-1.4885315218\times 10^9$ & $-2.2176133068\times 10^9$ & $-5.817819204726151\times 10^{11}$ & $6.239063727604102\times 10^{11}$\\
12 & $-7.703165634\times 10^{8}$& $-1.1555006405\times 10^9$ & $-2.971394997875579\times 10^{11}$ & $3.963717204418597\times 10^{11}$\\
14 & $-2.097322521\times 10^8$ & $2.5803602\times 10^6$ & $1.845822921481368\times 10^{11}$ & $7.077668090369869\times 10^{10}$\\
16 & $-3.04891722\times 10^7$ & $8.81906969\times 10^7$ & $1.694863702912681\times 10^{11}$ & $-1.205024034661577\times 10^{11}$\\
18 & $2.7037212\times 10^6$& $1.25437446\times 10^7$ & $-4.145362988742600\times 10^{10}$ & $-1.433538741513498\times 10^{11}$\\
20 & $3.3458376\times 10^6$& $-4.8988704\times 10^6$ & $-8.245022445306935\times 10^{10}$ & $-7.489713099273302\times 10^{10}$\\
22 & $1.187265808301358\times 10^6$ & $-1.666084222225726\times 10^6$ & $-7.660268981289846\times 10^8$ & $-1.561231898093660\times 10^{9}$\\
24 & $2.470076374928693\times10^5$ & $1.317431893346992\times 10^5$ & $3.538908387149367\times 10^{10}$& $3.358514735638041\times 10^{10}$\\
26 & $1.550608890427557\times 10^4$& $1.463028433008023\times 10^5$ & $8.354220074961258\times 10^ 9$ & $3.149037995362013\times 10^{10}$\\
28 & $-1.163077617533653\times 10^4$ & $1.121638684708260\times 10^4$ & $-1.336914484470230\times 10^{10}$ & $1.314270726948931\times 10^{10}$\\
30 & $-5.949362622274843\times 10^3 $& $-9.550578362817516\times 10^3$ & $-6.639667436677997\times 10^9$ & $-2.683420483299866\times 10^{9}$\\
32 & $-1.613595912491654\times 10^3$ & $-2.249168639875042\times 10^3$ & $4.258273988089967\times 10^9$ & $-8.603193097267729\times 10^9$ \\
34 & $-2.333879944573368\times 10^2$ & $4.033013968161846\times 10^2$ & $3.832168731738409\times 10^9$ & $-6.627186370824881\times 10^9$ \\
36 & $2.133138127038934\times 10^1$ & $2.311243264675432\times 10^2$ & $-9.654119269402733\times 10^8$ & $-2.049870598736127\times 10^9$ \\
38 & $2.585336000530348\times 10^1$ & $2.807625658141711$ & $-1.872722773415759\times 10^9$ & $1.220345600570677\times 10^9$ \\
40 & $9.132066749772124$ & $-1.732110087019092\times 10^1$ & $-9.661717995050006\times 10^6$ & $2.080653450746944\times 10^9$ \\
 \noalign{\hrule height 1.5pt}
\end{tabular}

\end{center}
}
\end{sidewaystable}
\begin{table}
\caption{Errors in the even ($\ell=2,4,6...$) zonal ($m=0$)
normalized  Stokes coefficients $\overline{C}_{\ell 0}=C_{\ell
0}/\sqrt{2\ell +1}$ for various Earth gravity models up to degree
$\ell=40$. They are not the formal statistic errors but have been
calibrated, although tentatively. Recall that $J_{\ell}\equiv
-C_{\ell 0}$. The references for the chosen Earth gravity models
solutions are: EIGEN-CG03C (F\"{o}rste et al. 2005);
EIGEN-GRACE02S (Reigber et al. 2005a); EIGEN-CG01C (Reigber et al.
2006); GGM02S (Tapley et al. 2005). } \label{geopot_models}
\begin{center}
\begin{tabular}{lllll}
\noalign{\hrule height 1.5pt}
 $\ell$ & EIGEN-CG03C & EIGEN-GRACE02S & EIGEN-CG01C & GGM02S\\
\hline
2 & 2.341$\times 10^{-11}$ & $5.304\times 10^{-11}$ & $3.750\times 10^{-11}$ & $1.1\times 10^{-10}$\\
4 & $3.778\times 10^{-12}$ & $3.921\times 10^{-12}$ & $6.242\times 10^{-12}$ & $8.3\times 10^{-12}$\\
6 & $1.840\times 10^{-12}$ & $2.049\times 10^{-12}$ & $2.820\times 10^{-12}$ & $4.5\times 10^{-12}$\\
8 & $1.170\times 10^{-12}$ & $1.479\times 10^{-12}$ & $1.792\times 10^{-12}$ & $2.8\times 10^{-12}$\\
10 & $8.576\times 10^{-13}$ & $2.101\times 10^{-12}$ & $1.317\times 10^{-12}$& $2.0\times 10^{-12}$\\
12 & $6.847\times 10^{-13}$ & $1.228\times 10^{-12}$ & $1.053\times 10^{-12}$& $1.8\times 10^{-12}$\\
14 & $5.806\times 10^{-13}$ & $1.202\times 10^{-12}$ & $8.931\times 10^{-13}$& $1.6\times 10^{-12}$\\
16 & $5.130\times 10^{-13}$ & $9.945\times 10^{-13}$ & $7.905\times 10^{-13}$& $1.6\times 10^{-12}$\\
18 & $4.684\times 10^{-13}$ & $9.984\times 10^{-13}$ & $7.236\times 10^{-13}$& $1.6\times 10^{-12}$\\
20 & $4.392\times 10^{-13}$ & $1.081\times 10^{-12}$ & $6.784\times 10^{-13}$& $1.6\times 10^{-12}$\\
22 & $4.624\times 10{-13}$  & $1.026\times 10{-12}$  & $7.152\times 10{-13}$ & $1.6\times 10^{-12}$ \\
24 & $4.912\times 10{-13}$  & $9.945\times 10{-13}$  & $7.600\times 10{-13}$ & $1.7\times 10^{-12}$ \\
26 & $5.260\times 10{-13}$  & $1.067\times 10{-12}$  & $8.148\times 10{-13}$ & $1.7\times 10^{-12}$ \\
28 & $5.664\times 10{-13}$  & $1.150\times 10{-12}$  & $8.784\times 10{-13}$ & $1.8\times 10^{-12}$ \\
30 & $6.140\times 10{-13}$  & $1.248\times 10{-12}$  & $9.528\times 10{-13}$ & $1.9\times 10^{-12}$ \\
32 & $6.684\times 10^{-13}$ & $1.359\times 10^{-12}$ & $1.038\times 10^{-12}$ & $1.9\times 10^{-12}$\\
34 & $7.312\times 10^{-13}$ & $1.488\times 10^{-12}$ & $1.136\times 10^{-12}$ & $2.1\times 10^{-12}$ \\
36 & $8.028\times 10^{-13}$ & $1.635\times 10^{-12}$ & $1.248\times 10^{-12}$ & $2.4\times 10^{-12}$ \\
38 & $8.852\times 10^{-13}$ & $1.803\times 10^{-12}$ & $1.376\times 10^{-12}$ & $2.5\times 10^{-12}$ \\
40 & $9.784\times 10^{-13}$ & $1.995\times 10^{-12}$ & $1.523\times 10^{-12}$ & $2.5\times 10^{-12}$ \\
 \noalign{\hrule height 1.5pt}
\end{tabular}
\end{center}
\end{table}
\begin{table}
\caption{Mismodelled combined classical precessions
$\left|\left(\dot\Omega_{.\ell}^{\rm
LAGEOS}+k_1\dot\Omega_{.\ell}^{\rm LAGEOS\
II}+k_2\dot\Omega_{.\ell}^{\rm Ajisai}+k_3\dot\Omega_{.\ell}^{\rm
Jason-1}\right)\right|\delta J_{\ell}$, in mas yr$^{-1}$, for
various Earth gravity models up to degree $\ell=40$. The results
of Table \ref{prec_coef} and Table \ref{geopot_models} have been
used. The percent upper bound of the total error $\delta\mu_{\rm
LT}$, for a given model, is obtained by summing all the elements
of the column corresponding to the model. The result is reported
in the last line. The predicted Lense-Thirring slope is 49.5 mas
yr$^{-1}$. } \label{comb_err}
\begin{center}
\begin{tabular}{lllll}
\noalign{\hrule height 1.5pt}
$\ell$ & EIGEN-CG03C & EIGEN-GRACE02S & EIGEN-CG01C & GGM02S\\
\hline
2 &- &- &- &- \\
4 & -& -& -& -\\
6 & -&- & -&- \\
8 & 0.141 & 0.178& 0.216& 0.338\\
10 & 0.170 & 0.417 & 0.261& 0.397\\
12 & 0.093& 0.168 & 0.014& 0.246\\
14 & 0.011& 0.023 & 0.017& 0.031\\
16 & 0.027& 0.052 & 0.041& 0.084\\
18 & 0.027& 0.058 & 0.042& 0.092\\
20 & 0.013& 0.032 & 0.020& 0.047\\
22 & $\mathcal{O}(10^{-4})$ & $\mathcal{O}(10^{-4})$ & $\mathcal{O}(10^{-4})$ & $\mathcal{O}(10^{-4})$\\
24 & 0.007 & 0.014 & 0.011& 0.025\\
26 & 0.008& 0.016& 0.012& 0.026\\
28 & 0.004 & 0.008 & 0.006& 0.013\\
30 & $\mathcal{O}(10^{-4})$ & 0.001 & 0.001& 0.003\\
32 & 0.003 & 0.006 & 0.005 & 0.009\\
34 & 0.002 & 0.005 & 0.004 & 0.008\\
36 & $\mathcal{O}(10^{-4})$ & 0.002 & 0.001 & 0.003\\
38 & $\mathcal{O}(10^{-4})$ & 0.001 & 0.001 & 0.002\\
40 & 0.001 & 0.002 & 0.002 & 0.003\\
 \hline
$\delta\mu_{\rm LT}\ (\%)$& 1.0 & 1.9 & 1.3 & 2.6\\
 \noalign{\hrule height 1.5pt}
\end{tabular}
\end{center}
\end{table}
\begin{table}
\caption{Error budget. The measurement error goes as 3$\%/N$,
where $N$ is the number of years of the observational time span:
we assume a $\sim 1$ m rms error over that time interval. A
duration of at least 3 years is required to average out the $K_1$
solar tide perturbation on the LAGEOS node whose period, equal to
that of the satellite's node, amounts to 2.84 years. The other
nodes have shorter periods. The upper limit for the
non-gravitational perturbation is over 2 years.} \label{tabwag}
\begin{center}
\begin{tabular}{llll}
\noalign{\hrule height 1.5pt}
Type of error & geopotential & measurement & non-gravitational \\
\hline
& 1$\%$ & 3$\%$ & 4$\%$\\
 \noalign{\hrule height 1.5pt}
\end{tabular}
\end{center}
\end{table}
%
%

\end{document}